\documentclass[prx,superscriptaddress,twocolumn,showpacs,preprintnumbers,floatfix]{revtex4}
\usepackage{amsmath,amssymb}
\usepackage{graphicx}
\usepackage[colorlinks, citecolor=blue]{hyperref}

\begin{document}

\title{Parity Anomaly of Lattice Maxwell Fermions in Two Spatial Dimensions}
\author{Jiang Zhou}
\affiliation{Department of Physics, Guizhou University, Guiyang 550025, PR China}
\author{Xiaosi Qi}
\affiliation{Department of Physics, Guizhou University, Guiyang 550025, PR China}
\author{Yajie Wu}
\affiliation{School of Science, Xi'an Technological University, Xi'an 710032, PR China}
\author{Su-Peng Kou}
\affiliation{Center for Advanced Quantum Studies, Department of Physics, Beijing Normal
University, Beijing 100875, China}
\date{\today}

\begin{abstract}
Unconventional lattice fermions with high degeneracies beyond Weyl and Dirac fermions have attracted intensive attention in recent years. In this paper, attention is drawn to the pseudospin-1 Maxwell fermions and the $(2+1)$ dimensional parity anomaly, which goes beyond the scope of "fermion doubling theorem". We have derived the Hall conductivity of a single Maxwell fermion, and showcased each Maxwell fermion contributes a quantized Hall conductance $e^{2}/h$. We observe that the parity is spontaneously broken in the effective theory of lattice Maxwell fermions interacting with an (auxiliary) $U(1)$ gauge field, leading to an effective anomaly-induced
Chern-Simons theory.  An interesting observation from the parity anomaly is that the lattice Maxwell fermions beyond the "fermion doubling theorem", so there exists single Maxwell fermion in a lattice model. In
addition, our work indicates the quantum anomaly in odd-dimensional spinor space.
\end{abstract}

\maketitle

\section{Introduction}

Dirac and Weyl fermions are massless relativistic particles predicted
firstly in the context of high-energy physics. In the standard model of
particle physics, they describe the matter fields. In condensed matter
physics, since the low-energy excitations around some band degenerate points
exhibit a relativistic dispersion relation, they can be regarded as emergent
Dirac or Weyl fermions. In recent years, these emergent relativistic
particles have attracted great attention.

Unlike the relativistic fermions in quantum field theory, the low-energy
excitations in solids, dubbed as pseudospin fermions, refers to pseudospin
that depends on the geometry of lattice. The fermions in solid preserve
space group rather than Poincare group, so rich fermions may exist in solid
universe. Started from the discovery of the celebrated Dirac fermions in
graphene, the unconventional quasiparticle that have no analog in quantum
field theory have stimulated an enormous interest\cite
{uf1,uf2,uf3,uf4,uf5,uf6,uf7,uf8,uf9,uf10,uf11,ph1,ph2,thrtr}. Recently, a
new type of fermion in three dimensions was predicted in topological
semimetal called unconventional fermions with highly degenerate points\cite
{trip,uf}, for example the triply degenerate fermions in molybdenum phosphide
material\cite{trip} and Maxwell fermions\cite{uf5}. The Maxwell fermions are triply degenerate fermionic excitations with pesudospin-1, the energy disperse linearly
near the triply degenerate points. As suggested by its name, they are described by the Hamiltonian which is analogous to massless spin-1 photons. Many literatures have study the triply degenerate fermion in various two- and three- dimensional families\cite{uf8,uf9,ph1,ph2,thrtr} and cold-atom optical lattice\cite {uf5,uf6,uf7}.
The unconventional fermions
show some extraordinary properties with respect
to graphene fermion, such as super-Klein tunneling effect\cite
{uf2,uf8,ph1} and bosonic topological phases\cite{ph2}. Moreover, the pairing
between higher spin fermions was suggested to explain the unconventional
superconducting of half-Heusler compounds\cite{scsm4}. For example, the
strong spin-orbit coupled half-Heusler compounds, in particular YPtBi and
LuPtBi, are described by $3/2$ spin-vector-coupling, the Cooper pair between
spin-$3/2$ fermions yields the pairing term for superconductor\cite%
{scsm1,scsm2,scsm3}.

Anomaly is a fundamental phenomenon for Dirac or Weyl fermions in condensed matter systems and it leads to some important physical consequences\cite{anomaly,anomaly2}. The $(3+1)$-D
axial (chiral) anomaly closely relates to the topological magneto-electric
effect of topological insulator\cite{qxl2,chi5,anomaly3}. On the contrary,
it relates to chiral magnetic effect and strong CP problem in quantum
chromodynamics\cite{chi2}. The $(2+1)$-D parity anomaly has close relation
with quantum Hall effect\cite{pan,pan2}. The $(1+1)$-D chiral anomaly will
lead to fractional charged soliton\cite{soli}. All these anomaly arise from
the breaking of the symmetry by quantum effect and the background topology
of the gauge field.

As we all known, there is a Nielsen-Ninomiya\cite{nn} theorem or "fermion doubling" theorem
which states that the number of chiral fermions on any lattice must be even, i.e., there are two massive Dirac fermions in Haldane model. In Qi-Wu-Zhang model for Chern insulator\cite{qwz}, the Nielsen-Ninomiya theorem guarantees there is another Dirac fermion with mass $-m$.
For triply degenerate lattice Maxwell fermions, however, a question then
arises: Does the Maxwell fermion with a definitive helicity in lattice strictly satisfies the "fermion doubling theorem"?
So far, whether the theory of lattice Maxwell fermions interacting with a gauge
field exhibits anomaly or not remains unclear. If yes, are there any new
physical implications? To answer these two important questions, in this
paper, we focus our attention on the parity anomaly of the Maxwell fermions
in two dimensions and discuss its physical consequences.

The remainder of this paper is organized as follows: in Sec.II we present the
Maxwell fermions in Lieb lattice and describe its low-energy Hamiltonian.
In Sec.III, we solve the Landau levels and calculate the Hall conductivity of single Maxwell fermion.
In Sec.IV, we study the effective theory of the Maxwell fermions coupled with the electromagnetic
field (Abelian $U(1)$ gauge field) upon integrating over the fermionic field, we obtain the effective theory
as Chern-Simons theory, so the parity is broken. Some physical implications for the fermion doubling theorem is discussed in this section.
Finally, the summary and some discussions are given in Sec.V.

\section{Lattice Maxwell fermions in the Lieb Lattice}

The lattice Maxwell fermions are fermionic particle with pseudospin-1 emerged
from latticel\cite{uf5}, which is analogous to the massless Weyl fermions
with spin-1/2. In quantum field theory, the particles with integral spins
can not be fermions. However, the fermions in lattice systems preserve only
some discrete space groups, so it's potential to find pseudospin-1 fermionic
excitations in condensed matter. Generally, the corresponding Bloch
Hamiltonian for Maxwell fermions can be written as
\begin{equation}  \label{e1}
H(\mathbf{k})=\mathbf{d}(\mathbf{k})\cdot \mathbf{S}
\end{equation}%
where $\mathbf{S}$ satisfies $[S_{i},S_{j}]=i\epsilon _{ijk}S_{k}$, $\mathbf{%
S}^{2}=S(S+1)$ with $S=1$, $\mathbf{d}(\mathbf{k})=[d_{x}(\mathbf{k}),d_{y}(%
\mathbf{k}),d_{z}(\mathbf{k})]$ is the Bloch vectors. There are two
different schemes to realize Maxwell fermions, which can be reached by using
three-component fermionic (three internal degrees of freedom atoms) in
simplest lattice, or by using singlet-component fermionic atoms in three
sublattices systems.

\begin{figure}[tbp]
\centering
\scalebox{1.05}{\includegraphics*{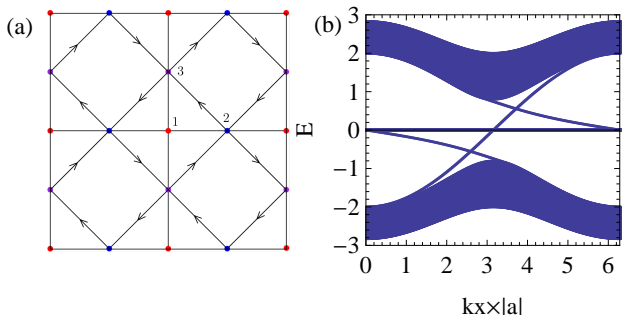}}
\caption{(Color online) (a)The sketch of Lieb lattice, an unit cell consists
of three sublattices denoted by 1, 2, 3. A electron acquires a factor $i$
when traveling along the arrow. (b) Illustration of the edge states, $%
\protect\lambda =0.2t$ in this plot.}
\label{fig2}
\end{figure}

The Lieb lattice (Fig.~\ref{fig2}(a)) provides a natural three sublattices
system. In analogy with the Haldane model, the Lieb lattice exhibits quantum
anomalous Hall effect with an additional phase dependent next
nearest-neighbor (NNN) hopping\cite{lieb0}. Now, the tight-binding model is given by
\begin{equation}
H=-t\sum_{\langle ij\rangle }c_{j}^{\dag }c_{i}+i\lambda \sum_{\langle
\langle ij\rangle \rangle }(\mathbf{\hat{v}}_{i}\times \mathbf{\hat{v}}%
_{j})_{z}c_{j}^{\dag }c_{i},
\end{equation}%
where $\langle ij\rangle $ denotes the summation over all nearest-neighbor
pairs with hopping amplitude $t$, $c_{i}^{\dag }(c_{i})$ are the
creation(annihilation) operators on site $i$. $\lambda $ plays the
role of NNN hopping (see Fig.\ref{fig2}(a)), and $(\mathbf{\hat{v}}
_{i}\times \mathbf{\hat{v}}_{j})_{z}=\pm 1$, $\mathbf{\hat{v}}_{i}$ and $%
\mathbf{\hat{v}}_{j}$ are the unit vectors along the two bonds constituting
the NNN. This model preserves translational but breaks time-reversal
symmetry. In terms of the momentum basis $\psi _{\mathbf{k}}=(c_{1\mathbf{k}%
},c_{2\mathbf{k}},c_{3\mathbf{k}})^{\text{T}}$, the Hamiltonian can be
rewritten as $H=\sum_{k}\psi _{\mathbf{k}}^{\dag }H(\mathbf{k})\psi _{%
\mathbf{k}}^{\dag }$ , where the Bloch Hamiltonian takes the form of Eq.(\ref%
{e1}) and $d_{x}(\mathbf{k})=-2t\cos k_{x}$, $d_{y}(\mathbf{k})=-2t\cos k_{y}
$, $d_{z}(\mathbf{k})=4\lambda \sin k_{x}\sin k_{y}$. Here the spin matrices
are given as
\begin{align}
& S_{x}=\left(
\begin{array}{ccc}
0 & 1 & 0 \\
1 & 0 & 0 \\
0 & 0 & 0%
\end{array}%
\right) ,S_{y}=\left(
\begin{array}{ccc}
0 & 0 & 1 \\
0 & 0 & 0 \\
1 & 0 & 0%
\end{array}%
\right) ,  \notag \\
& S_{z}=\left(
\begin{array}{ccc}
0 & 0 & 0 \\
0 & 0 & -i \\
0 & i & 0%
\end{array}%
\right) .
\end{align}%
These matrices form the specific foundation representation of \textrm{SU(3)}
group, named Gell-Mann matrices. Note that these matrices satisfy the
\textrm{SU(2)} algebra relation but do not form a Clifford algebra. For $%
\lambda =0$, the single particle spectrum consists of two dispersive bands
and one zero energy flat band
\begin{equation}
E_{\pm }=\pm 2t\sqrt{\cos ^{2}k_{x}+\cos ^{2}k_{y}},\text{ }E_{0}=0.
\label{dis}
\end{equation}
The three bands touch at the triply degenerate fermion point $\mathbf{K}_{F}=(\pi
/2,\pi /2)$, In the vicinity of the triply degenerate point, the effective
Hamiltonian has the form
\begin{equation}
  H(\mathbf{q})=v_Fq_{x}S_{x}+v_Fq_{y}S_{y},
\end{equation}
where $v_F=2t$ is the fermion velocity and $\mathbf{q=k-K}_{F}$. This effective
Hamiltonian describes a relativistic massless fermionic excitations,
so-called Maxwell fermions\cite{uf5}. We also note that there is a single
triply degenerate point in the reduced Brillouin zone of the Lieb lattice.

The system is of Chern insulating phase when $\lambda >0$, and the topological invariant of the lowest occupied band   is given by the first Chern number. For the lowest occupied band, we find
\begin{equation}
C_{1}=\frac{1}{2\pi }\int_{0}^{\pi }dk_{x}\int_{0}^{\pi }dk_{y}\mathbf{F}%
_{xy}=1,
\end{equation}%
where the Berry curvature $\mathbf{F}_{xy}=\mathbf{\nabla \times A(k)}$, and
the Berry connection is defined by the wave function of the lowest band $%
\mathbf{A(k)=-}i\langle \psi (\mathbf{k})|\mathbf{\nabla }_{\mathbf{k}}|\psi
(\mathbf{k})\rangle $. An important property of the topological nontrivial
states is the presence of spatially localized edge states. In order to
visualize the edge states, a standard method is to diagonalize the lattice
model with periodic boundary conditions imposed along one of the spatial
directions, say, $x$-direction. The energy spectrum as a function of
momentum are shown in Fig.~\ref{fig2} (b). We observe that there is only one
chiral edge state, and another branch of twisted edge state is interesting.
The single chiral reflects the bulk-edge correspondence of topological
states. Further, the chiral edge state can be studied analytically, this has
been done in Ref.~\onlinecite{cs1}.
\section{The Properties of Single Maxwell Fermion}
The parity anomaly for the Dirac fermions theory in $2+1$ dimensions is closely relevant to the quantum Hall effect of the fermions. For example, the anomaly will lead to a Chern-Simons term whose coefficient is the same as the value of Hall conductivity of the fermions. So, before study the parity anomaly of the Maxwell fermions, let's study some properties (Landau levels and Hall conductivity) of a single Maxwell fermion for comparison.
\subsection{Landau levels for Maxwell fermions}

\begin{figure}[tbp]
\centering
\scalebox{1.0}{\includegraphics*{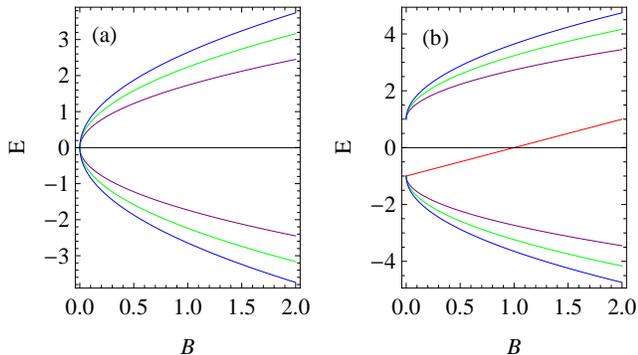}}
\caption{(Color online) The Landau-level of TDRFs versus magnetic field $B$.
(a) for massless fermion and (b) for massive fermions. In the massive case
(b), the red level shows the zeroth Landau level.}
\label{fig1}
\end{figure}

Consider massive Maxwell fermion with the Hamiltonian reads
\begin{equation}\label{massive}
  H(\mathbf{q})=v_Fq_{x}S_{x}+v_Fq_{y}S_{y}+mS_z,
\end{equation}
the incorporation of perpendicular magnetic field $\mathbf{B}=B\hat{z}$ can be implemented by replacing the
momentum with $\mathbf{q}\rightarrow \Pi =\mathbf{q}-e\mathbf{A}$,
where $\mathbf{A=(-}By,0,0\mathbf{)}$  is the Landau gauge. The procedure for a uniform magnetic
field lead to Peierls substitution on the lattice. One can prove that $%
[\Pi_{x},\Pi _{y}]=-i\hbar ^{2}/l_{B}^{2}$ with $l_{B}=\sqrt{\hbar /eB}$ to
be the magnetic length. To solve the energy level, it's convenient to
introduce a pair of ladder operators $a=\frac{l_{B}}{\sqrt{2}\hbar }(\Pi
_{x}-i\Pi_{y})$ and $a^{\dag }=\frac{l_{B}}{\sqrt{2}\hbar }(\Pi _{x}+i\Pi
_{y})$. Now, the Hamiltonian becomes
\begin{equation}
H=\frac{\hbar v_{F}}{\sqrt{2}l_{B}}(a^{\dag }S_{-}+aS_{+})+mS_{z},
\end{equation}
where $S_{\pm}=S_{x}\pm iS_{y}$. Without $M$, the Landau levels are
found to be
\begin{equation}
E_{n\pm }=\pm \sqrt{2n+1}\hbar v_{F}/l_{B},\text{ }n\geq 1,  \label{LLs}
\end{equation}
which is in contrast to the case of Dirac fermion $E_{n\pm }\sim\sqrt{2n}$.
There is a highly degenerate flat band lying exactly at $E_{n}=0$,
this is right the zero energy flat band under zero magnetic field and does
not contribute the Hall conductance. In addition, another zero energy level
becomes nontrivial, of which the wave function is $\psi
_{0}=(0,\phi_{0},-i\phi _{0})^{\text{T}}$, where $\phi _{0}$ satisfies $a\phi _{0}=0$.
To study the nontrivial zeroth energy level,
we consider a momentum dependence mass term $m\rightarrow m(k)=m-(\Pi
_{x}^{2}+\Pi_{y}^{2}) $, in terms of ladder operators, it reads
\begin{equation}
m(k)=m-2\hbar ^{2}/l_{B}^{2}(a^{\dag }a+1/2)].
\end{equation}
In this case, we find the energy spectrum for $n\geq 1$ come
into pairs, as depicted in Fig~\ref{fig1}(b). Special care need to be taken
for $n=0$ in the massive case, which yields
\begin{equation}
E_{n=0}=-m+\hbar ^{2}/l_{B}^{2}.
\end{equation}
Because the zero-th Landau level depends on the magnetic field and increases linearly as the magnetic
field, it crosses Fermi energy. Therefore, the zero-th Landau level provides
an one-dimensional chiral mode, which is hallmark of the nontrivial
topological edge states. Taking the continuum limit, the one-dimensional
chiral mode can be effectively described by the Lagrangian of free fermion:
\begin{equation}
\mathcal{L}=\frac{i}{2}\int dx\psi (\partial _{t}+\partial _{x})\psi ,
\end{equation}
where $\psi $ denotes single fermion field. From this free Lagrangian, under parity transformation, we observe the
chiral mode leads to violation of parity, which implies the parity anomaly in a topological nontrivial phase.
\subsection{Hall conductivity}
In this section, we calculate the Hall response of a single Maxwell fermion. The low-energy theory is described
by linear cone, which allows us to study it analytically. In general, the
dc response to external field can be obtained in terms of standard Kubo
formula\cite{qxl2}:
\begin{eqnarray}
\sigma _{ij}(\omega ) &=&\lim_{\omega \rightarrow 0}\frac{i}{\omega }
Q_{ij}(i\omega _{a}\rightarrow \omega +i0^{+}), \\
Q_{ij}(i\nu _{b}) &=&\frac{1}{V\beta }\sum_{\mathbf{k,}a}\text{Tr}[J_{i}(
\mathbf{k})\times  \notag \\
&&G(\mathbf{k},i(\omega _{a}+\nu _{b}))J_{j}(\mathbf{k})G(\mathbf{k},i\omega
_{a})],
\end{eqnarray}
where Tr means trace over spinor indices, the inverse
temperature $1/T=\beta$ and $V$ is the area of the system. The dc current is $J_{i}(
\mathbf{k})=\partial h(\mathbf{k})/\partial \mathbf{k}_{i}$, with $i=x,y$.
The single-body Green function is given by $G^{-1}(\mathbf{k}
,i\omega_{a})=i\omega _{a}-H(\mathbf{k})$, and $\omega _{a}=(2n+1)\pi /\beta
$ is Matsubara frequency. For the Massive Maxwell fermion in Eq.(\ref{massive}), we have
the current-current corrections
\begin{eqnarray}\label{polarization}
Q_{xy}(i\nu _{b}) &=&\frac{1}{V\beta }\sum_{\mathbf{k,}a}\text{Tr}\left[S_{x}
\frac{1}{i(\omega _{a}+\nu _{b})-H(\mathbf{k})}\right.  \notag \\
&&\left.\times S_{y}\frac{1}{i\omega _{a}-H(\mathbf{k})}\right].
\end{eqnarray}
Since the three representation matrices for spin $S=1$ do not
satisfy Clifford algebra, the best way to proceed is finding out the single-body Green function
explicitly. The Hall conductivities is obtained in the dc limit, say $\nu \sim 0$.
Finally, we obtain the result
\begin{eqnarray}\label{conductivity}
\sigma^{M}_{xy}&=&i\int \frac{d\omega d^{2}\mathbf{k}}{(2\pi )^{3}}\frac{\omega ^{2}m+(
\mathbf{k}^{2}+M^{2})m}{\omega ^{2}[\omega ^{2}-(\mathbf{k}^{2}+m^{2})]^{2}}
\notag \\
&=&-\frac{1}{2\pi }\frac{m}{|m|}.
\end{eqnarray}
More details about this calculation are presented in Appendix A. Putting
back the standard conductivity units, we have $\sigma^{M}_{xy}=-$sgn($m$)$
\times e^{2}/h$, which differs from the result of single Dirac cone $\sigma^{D}_{xy}=-$sgn($m$)$
\times e^{2}/2h$.
Here and below in this paper, we set the units $e=\hbar =1$ such
that the Hall conductivity in units of $e^{2}/h=1/2\pi $.
\section{Parity anomaly and implications}

In this section, let's discuss the parity of the Maxwell fermions. For some gauge theory, i.e. the Abelian quantum electrodynamics (QED) and the non-Abelian Yang-Mills theory, the quantum anomaly has deep connections with the topology of the background field.
In usual $(2+1)$-dimensions QED, parity (or
time-reversal) anomaly is great importance because it is relevant to the
quantized anomalous Hall effect of the fermions\cite{pan}. We shall derive the parity anomaly of the theory for
Maxwell fermions coupled to \textrm{U(1)} gauge fields (a special QED-type
theory). As far as we are aware, our work provides the first derivation of anomaly from a
Hamiltonian in odd-dimensional spinor space in $(2+1)$-dimensions.

\subsection{Invariance of the massless theory}

To demonstrate the parity anomaly more clearer, we begin by briefly
reviewing the anomaly in usual QED theory. The parity anomaly has been firstly
discussed by Redlich\cite{anomaly2} that arises from the derivation of
effective action upon integrating out the fermion fields. The action for
massless Dirac fermions coupled to \textrm{U(1)} gauge fields in three time-space
is invariant under gauge transformations and parity reflections. The gauge
invariance and parity invariance of effective action, however, depend on the
procedure used to regulate the ultraviolet divergences. There are two ways
to regulate the theory, we can regulate in a way that maintains parity well
but the effective action is not gauge invariant. Alternatively, we can use a
gauge invariant regularization at the expense of parity invariant (as
similar as Pauli-Villars scheme). The Pauli-Villars regulator field includes
a heavy mass which is not parity invariant in $(2+1)$-dimensions.

In order to preserve the gauge invariance, the Pauli-Villars regularization scheme amounts
to the introduction of another fermion with heavy mass $M$, then the regulated effective action is
\begin{equation}
I_{\text{eff}}^{R}[A]=I_{\text{eff}}[A,m=0]-I_{\text{eff}}[A,m\rightarrow
\infty ].  \label{e45}
\end{equation}
The second term produces a finite parity-violating topological action $%
\pm\pi W[A]$, that is the Chem-Simons (CS) secondary characteristic class%
\cite{anomaly2}. The gauge non-invariance of Chern-Simons action cancels the
gauge noninvariance of $I_{\text{eff}}[A]$ under a homotopically nontrivial
gauge transformation with winding number $n$, which yields a gauge invariant
but parity-violating action $I_{\text{eff}}^{R}[A]$.

In the same way, we are going to show the Maxwell fermions coupled to \textrm{U(1)}
gauge fields exhibits parity anomaly. The Lagrangian for free Maxwell fermion is given by
\begin{equation}
\mathcal{L}=\bar{\psi}(i\gamma ^{i}\partial _{i}-\zeta _{0}\beta ^{0}m)\psi
\end{equation}
where $\psi $ is a three-component spinor. Here, we use Minkowski space
metric $\eta^{\mu \nu }=\text{diag}(1,-1,-1)$
with $\mu $, $\nu =0,1,2$. To defined the adjoint spinor, we have introduced the matrix $\zeta _{0}=$
diag$(-1,1,1)$ such
that the adjoint spinor $\bar{\psi}=\psi ^{\dag }\zeta _{0}$,. The gamma matrices are defined as $\gamma^{1}=\zeta _{0}S_{x}$, $
\gamma ^{2}=\zeta _{0}S_{y}$, $\gamma^{0}=\zeta _{0}$, $\beta ^{0}=S_{z}$.
With the help of Legendre transformation, one can check above Lagrangian recovers the standard Hamiltonian of Maxwell fermions, see Eq.(\ref{massive}) for example. It seems that the Lagrangian unlike the
usual form of Dirac fermions, this arises from the absence of
three-dimensional matrix representation of anti-commutative Clifford algebra.
The discrete parity transformation in $\left( 2+1\right) $-dimensions is
unusual. In two-dimensional plane, the parity transformation acts as $
(x,y)\rightarrow (x,-y)$ rather than $(x,y)\rightarrow (-x,-y)$. The latter
is equivalent to Lorentz rotation $\Lambda_L$ with $\det (\Lambda_L)=1$. Under parity $P$
, the fermion fields transform as $P\psi P^{-1}=\Lambda_P \psi$,
where the parity satisfies $\det ({\Lambda_P})=-1$. The gauge fields
transform as $P\vec{A}P^{-1}=(A_{0},A_{x},-A_{y})$. Therefore, one can check
that the massless QED theory
\begin{equation}
\mathcal{L}=\bar{\psi}(i\gamma ^{\mu }\partial _{\mu }-e\gamma ^{\mu }A_{\mu
})\psi,
\end{equation}
is parity invariant, where the parity satisfy $P\gamma^1P^{-1}=\gamma^1$, $P\gamma^2P^{-1}=-\gamma^2$,
$P\gamma^0P^{-1}=\gamma^0$.
We observe that, however, the mass term $m\psi
^{\dag}\beta ^{0}\psi $ and the CS term $\varepsilon ^{\mu \nu \lambda
}A_{\mu }\partial _{\nu }A_{\lambda }$ are not parity invariant. Instead,
they change sign under parity transformation. Moreover, one can show that the
special massless QED theory is time-reversal invariant but the mass term and
Chern-Simons term are not invariant.
\begin{table}[t]
\caption{The nonzero antisymmetric part $\Pi _{A}$ of polarization tensor}%
\begin{tabular}{lllll}
\hline\hline
& tensor\hspace{0.5in} & result \hspace{0.6in} & tensor\hspace{0.5in} & result \\ \hline
& $\Pi_A^{01}$ & $\frac{e^2}{2\pi}q_2$ & $\Pi_A^{10}$ & $-\frac{e^2}{2\pi}q_2$ \\
& $\Pi_A^{12}$ & $\frac{e^2}{2\pi}q_0$ & $\Pi_A^{21}$ & $-\frac{e^2}{2\pi}q_0$ \\
& $\Pi_A^{20}$ & $\frac{e^2}{2\pi}q_1$ & $\Pi_A^{02}$ & $-\frac{e^2}{2\pi}q_1$ \\ \hline\hline
\end{tabular}
\end{table}
\subsection{Perturbation calculations of Chern-Simons term}
We have show the invariance of the theory with Maxwell fermions coupled with a gauge field.. In this section, we
will demonstrate that the Chern-Simons action will arise in the effective
action of gauge fields after regulating the theory in a gauge invariant way.
To this end, we use Pauli-Villars regularization and regulate the theory
according to Eq.~(\ref{e45}).

Now, consider the massive fermions, the effective action can be derived by
integrating over the fermion fields in the functional integral:
\begin{eqnarray}
Z &=&\int d\bar{\psi}d\psi dA\exp \left( i\int d^{3}x\mathcal{L}\right) , \\
\mathcal{L} &=&\frac{1}{2}F_{\mu \nu }F^{\mu \nu }+\bar{\psi}(i\gamma ^{\mu
}D_{\mu }-\zeta_{0}\beta ^{0}m)\psi ,
\end{eqnarray}
where $D_{\mu }=\partial _{\mu }+ieA_{\mu }$ is the covariant derivative, $%
F^{\mu \nu }=\partial _{\mu }A_{\nu }-\partial _{\nu }A_{\mu }$ is the usual
notation of field strength. The effective action, upon integrating the
fermion fields, is given by
\begin{equation}
Z=\int dA\exp \left( i\int d^{3}x\frac{1}{2}F_{\mu \nu }F^{\mu \nu }+I_{
\text{eff}}[A,m]\right) ,
\end{equation}
where
\begin{equation}
I_{\text{eff}}[A,m]=-iN_{f}\ln (i\gamma ^{\mu }D_{\mu }-\zeta _{0}\beta
^{0}m).
\end{equation}
We have introduced the fermion flavors $N_{f}$ which will be important for
the gauge invariance of the action. Taking the large $m$ limit, this allows
us do perturbative expansion at one-loop order, all other higher-order
graphs are order $O(1/m)$ and thus vanish when $m\rightarrow \infty $. As
shown before\cite{anomaly2} , in the Abelian theory, only the vacuum polarization is ultraviolet divergent, therefore require regularization.

To seek the parity anomaly term, to order $e^{2}$, we have
\begin{equation}\label{eff}
I_{\text{eff}}[A,m]=-i\frac{N_{f}}{2}\int \frac{d^{3}q}{(2\pi )^{3}}A_{\mu
}(q)\Pi ^{\mu \nu }(q)A_{\nu }(-q),
\end{equation}
where the polarization tensor is given by
\begin{equation}
\Pi ^{\mu \nu }(q)=e^{2}\int \frac{d^{3}p}{(2\pi )^{3}}\text{Tr}\left[
\gamma ^{\mu }S(p+q)\gamma ^{\nu }S(p)\right] ,
\end{equation}
with $S(p)=i/(\gamma p-\zeta _{0}\beta ^{0}m)$. The polarization tensor is
superficially linearly divergent as the cut-off $\Lambda \rightarrow \infty $%
. Since our gamma matrices defined in odd-dimensions representation do not
have nice trace and anti-commutative properties as in even-dimensional
representation, to derive the last result, the possible nonzero tensor should be calculated one by one.
Here, we restrict our attention to $\Pi ^{01}(q)$, after performing matrix
inversion and trace operation, we obtain the antisymmetric part:
\begin{eqnarray}
\Pi _{A}^{01}(q,m)&=&ie^{2}\int \frac{d^{3}p}{(2\pi )^{3}}\frac{mq_{2}(\mathbf{p}
^{2}+m^{2}-p_{0}^{2})}{p_{0}^{2}[(p+q)^{2}-m^{2}][p^{2}-m^{2}]}  \notag \\
&=&-ie^{2}q_{2}(\Pi _{A1}+\Pi _{A2}),
\end{eqnarray}
\begin{eqnarray}
\Pi _{A1} &=&\int \frac{d^{3}p}{(2\pi )^{3}}\frac{m}{
[(p+q)^{2}-m^{2}][p^{2}-m^{2}]}  \notag \\
&=&\frac{i}{8\pi }\frac{m}{|m|},
\end{eqnarray}
\begin{eqnarray}
\Pi _{A2} &=&\int \frac{d^{3}p}{(2\pi )^{3}}\frac{-m(\mathbf{p}^{2}+m^{2})}{
p_{0}^{2}[(p+q)^{2}-m^{2}][p^{2}-m^{2}]}  \notag \\
&=&\frac{3i}{8\pi }\frac{m}{|m|}.
\end{eqnarray}
To reach the last result, we have (i) introduced Feynman parameter to
rewrite the denominator, (ii) shifted integration variable and taken the
limit $q_{2}\rightarrow 0$, (iii) carried out the integration by parts. The
calculations of polarization function can be referred to Appendix. For $
m\rightarrow \infty $, we find
\begin{equation}
\Pi _{A}^{01}(m\rightarrow \infty )=\frac{m}{|m|}\frac{e^{2}}{2\pi }
\varepsilon ^{012}q_{2}.
\end{equation}
Repeating the calculation for other nonzero tensors, the results are
displayed in Table I.

Putting these results together, we observe
\begin{equation}
\Pi _{A}^{\mu \nu }(m\rightarrow \infty )=\frac{m}{|m|}\frac{e^{2}}{2\pi }
\varepsilon ^{\mu \nu \lambda }q_{\lambda }.
\end{equation}
By Lorentz invariance, the symmetric part is linearly divergent. In general,
it must be of the form
\begin{equation}
\Pi _{S}^{\mu \nu }(m\rightarrow \infty )=\text{constant}\times e^{2}\Lambda
\eta ^{\mu \nu },
\end{equation}
where $\Lambda $ is a momentum cutoff. Inserting these results into Eq.(\ref{eff}) yields
\begin{eqnarray}
I_{\text{eff}}[A,m &\rightarrow &\infty ]=-i\frac{N_{f}}{2}(\text{constant}
\times e^{2}\Lambda A^{2}  \notag \\
&&-\frac{im}{|m|}\frac{e^{2}}{2\pi }\varepsilon ^{\mu \nu \lambda }A_{\mu
}\partial _{\nu }A_{\lambda }).
\end{eqnarray}
The divergent term in $I_{\text{eff}}[A,m=0]$ cancels the divergent term in $%
I_{\text{eff}}[A,m\rightarrow \infty ]$. We obtain the regularized and
finite effective action as
\begin{equation}\label{cs}
I_{\text{eff}}^{R}[A]=\text{finite}\pm \frac{N_{f}}{2}\frac{e^{2}}{2\pi }
\varepsilon ^{\mu \nu \lambda }A_{\mu }\partial _{\nu }A_{\lambda }.
\end{equation}
where the symbol $\pm $ depends on the sign of $m$. The finite part
conserves parity, and the parity violation comes from the Chern-Simons term.
In summary, we have shown that the parity is broken to
maintain the gauge invariance of the regulated effective action.
\subsection{Physical implications}
To see the physical implications, It's helpful to write the Chern-Simons action as a standard form
\begin{equation}\label{cs2}
\mathcal{L}_{CS}=\frac{k}{4\pi }\varepsilon ^{\mu \nu \lambda }A_{\mu
}\partial _{\nu }A_{\lambda },
\end{equation}
The Chern-Simons action is closely relevant to the topology of the gauge field, it does not gauge-invariant if the space-time manifold is open. Under a large gauge transformation, the action is invariant on a closed space-time manifold
unless the parameter $k$ is quantized as $k\in $ integer.

The parity anomaly predicts
\begin{align}
  k=\left\{\begin{array}{c}
        1/2, \quad \text{single Dirac fermion} \\
        1,\quad \text{single Maxwell fermion}
      \end{array}\right.
\end{align}
For Dirac fermions in usual QED theory, it is puzzling because the Chern-Simons term requires $
k $ should be an integer. The understanding of the puzzle lies in the
Nielsen-Ninomiya theorem or "fermion doubling theorem" which implies that the number of local
fermions on any lattice must be even. Three typical models are of great help for us: the
Haldane model for quantum anomalous Hall effect\cite{pan}, the Qi-Wu-Zhang model
for Chern insulator\cite{qwz}, and the axion topological insulator\cite{qxl2}. When the
time-reversal or parity is broken in Haldane model, there are
two Dirac fermions due to the "fermion doubling theorem"\cite{nn}, each of
which contributes a half quantized Hall conductivity such that the total
value equals to be an integer. In Qi-Wu-Zhang model, there exists a single
fermion in the Brillouin zone. In this case, we need consider a compact Brillouin zone
in lattice model and then the "fermion doubling theorem" plays the role of
Pauli-Villars regulator with mass $-m$. As a result, the linearly divergent
has canceled each other and the Chern-Simons term is doubled, yields integer
Chern number that depends on the sign of $m$. In axion topological
insulator, the bulk states preserve time-reversal but the boundary state is
time-reversal broken. So, there is a single massive fermion on each
boundary. The "fermion doubling theorem" is avoided due to the massive fermion
lives on different boundary. Therefore, the axion topological insulator
supports surface Hall effect with half Hall conductivity.

As shown in Eq.(\ref{cs}), the effective theory for Maxwell fermions coupled with gauge field is
\begin{equation}
I_{\text{eff}}^{R}[A]=\frac{N_{f}}{4\pi }\varepsilon ^{\mu \nu \lambda
}A_{\mu }\partial _{\nu }A_{\lambda }.
\end{equation}
The minimum value $N_{f}=1$ satisfies
the quantized condition, which implies that there exists single local
Maxwell fermion in discrete lattice. That is, the lattice Maxwell fermions go beyond the scope of
"fermion doubling theorem".
Furthermore, the Chern-Simons term predicts the Hall current
\begin{equation}
J^{u}=\frac{\delta I_{\text{eff}}^{R}[A]}{\delta A_{\mu }}=\frac{1}{2\pi }
\times \varepsilon ^{\mu \nu \lambda }\partial _{\nu }A_{\lambda },
\end{equation}
this result agrees with the Hall conductivity (or Chern number) in the Lieb lattice: $\sigma ^{xy}=1$
in units of $e^{2}/h$. It also agrees with the analytical results Eq.(\ref{conductivity}).

\section{Summary and discussions}

In summary, we have studied the parity anomaly of the lattice Maxwell fermions.
In the same way as usual QED theory, we find that the parity is spontaneous1y broken in the effective theory of Maxwell fermions coupled
with U(1) gauge field. We have derived the effective theory as
anomaly-induced Chern-Simons theory, as shown in Eq.(\ref{cs}), in which both
the parity and time-reversal are broken.
For the Maxwell fermions, the Chern-Simons action predicts a Hall conductivity of the ground states current with the value $e^{2}/h$, which agrees with the result obtained from framework of the Kubo formula. Moreover, there exist triply degenerate fermion systems with large quantized anomalous Hall conductivity, given by $\sigma_{xy}=2e^2/h$, and this have been studied in Refs.~[\onlinecite{uf4,uf5,thrtr,ph2}].

The gauge invariance of the Chern-Simons term [see Eq.(\ref{cs2})] on a closed manifold under large gauge transformations require the coefficient $k$ must be an integer. For Dirac fermions, the coefficient of the Chern-Simons term gives $k=1/2$, in this case, the "fermion doubling theorem" guarantees another Dirac fermion, yielding the coefficient with the value $k=1$ totally. However, for triply degenerate lattice Maxwell fermions, the minimum value of fermion flavors $N^{min}_f=1$ meets the requirement of gauge-invariant under large gauge transformations. So, there exists single Maxwell fermion in discrete lattice, which implies that the Maxwell fermions go beyond the scope of "fermion doubling theorem". Finally, our results also suggests anomaly can take place in odd-dimensional spinor space.

\begin{acknowledgments}
This work is supported by NSFC (Nos.11647111, 11604060, 11674062, 11974053),
and the work is supported partly by the Science and Technology Project and
Talent Team Plan of Guizhou province(2017-5610,2017-5788).
\end{acknowledgments}

\begin{appendix}
\section{Hall conductivity for single Maxwell fermion}
In this Appendix, we present some details of the Hall conductance for single Maxwell fermion. Starting from the Kubo formula, we need to calculate the explicit expression for the Green function since the gamma matrices in odd dimensions do not have "nice" trace property as in even dimensions. The Green function is given by
$G=[i\omega _{a}-H(\mathbf{k})]^{-1}$. Inserting the Green function into Eq.(\ref{polarization}), we then get the polarization function
\begin{widetext}
\begin{equation}
Q_{xy}(i\nu _{b})=\frac{i}{V\beta }\sum_{\mathbf{k,}a}\frac{i\omega
_{a}(i\nu _{b})^{2}m+(i\omega _{a})^{2}i\nu _{b}M+i\nu _{b}\mathbf{k}
^{2}m+i\nu _{b}m^{3}}{i\omega _{a}(i\omega _{a}+i\nu _{b})[(i\omega
_{a}+i\nu _{b})^{2}-(\mathbf{k}^{2}+m^{2})][(i\omega _{a})^{2}-(\mathbf{k}
^{2}+m^{2})]},
\end{equation}
where the identity Tr$(S_{x}S_{y}S_{z})=i$ has been used. Only the antisymmetric part survive and all terms that are symmetric for $x\leftrightarrow y$ had been neglected to reach this result. The Hall conductivities is obtained in the dc limit, say $\nu \sim 0$. After analytic continuation $i\nu _{n}\rightarrow \nu +i0^{+}$, and then taking the dc limit, we have
\begin{eqnarray}
\sigma _{xy}&=&\lim_{\nu \rightarrow 0}\frac{i}{\nu }Q_{ij}(\nu +i0^{+})
=i\int \frac{d\omega d^{2}\mathbf{k}}{(2\pi )^{3}}\frac{m}{[\omega ^{2}-(
\mathbf{k}^{2}+m^{2})]^{2}}
+i\int \frac{d\omega d^{2}\mathbf{k}}{(2\pi )^{3}}\frac{(\mathbf{k}
^{2}+m^{2})m}{\omega ^{2}[\omega ^{2}-(\mathbf{k}^{2}+m^{2})]^{2}}.
\end{eqnarray}
\end{widetext}
Performing Wick rotation $\omega =i\omega _{E}$, the first integral is elementary:
\begin{eqnarray}
i\int \frac{d\omega d^{2}\mathbf{k}}{(2\pi )^{3}}\frac{m}{[\omega ^{2}-(
\mathbf{k}^{2}+m^{2})]^{2}}
=-\frac{m}{8\pi |m|}
\end{eqnarray}
In the same way, for the second integral, we have
\begin{eqnarray}
i\int \frac{d\omega d^{2}\mathbf{k}}{(2\pi )^{3}}\frac{(\mathbf{k}
^{2}+m^{2})m}{\omega ^{2}[\omega ^{2}-(\mathbf{k}^{2}+m^{2})]^{2}}
=\frac{-3m}{8\pi |m|}.
\end{eqnarray}
In order to reach the last result, we have to integral over $\omega _{E}$ by parts many times. In summary, we have obtained the total Hall conductivity for single Maxwell fermion $\sigma^{M}_{xy}=\frac{-m}{2\pi|m|}$
\end{appendix}

\end{document}